\documentstyle[graphicx,psfrag]{aipproc2}

\renewcommand{\Re}{{\rm R}}

\newcommand{\spi}{i^o}
\newcommand{\scri}{{\cal I}}

\newcommand{\rad}{r_\Delta^{}}
\newcommand{\QD}{Q_\Delta^{}}
\newcommand{\PD}{P_\Delta^{}}
\newcommand{\SD}{S_\Delta^{}}
\newcommand{\MD}{M_\Delta^{}}
\newcommand{\HD}{H_\Delta^{}}
\newcommand{\HI}{H_\infty^{}}
\newcommand{\Ht}{H_t^{}}
\newcommand{\ONI}{\Omega_{\cal I}^{}}
\newcommand{\HNI}{H_{\cal I}^{}}

\begin{document}

\title  {A Hamiltonian Approach to the Mass of\\ Isolated Black Holes}
\author {Christopher Beetle and Stephen Fairhurst}
\address{Center for Gravitational Physics and Geometry\\
         Department of Physics, The Pennsylvania State University\\
         University Park, PA 16802}
\maketitle

\begin{abstract}
Boundary conditions defining a {\it non-rotating isolated horizon\/} are given
in Einstein--Maxwell theory.  A spacetime representing a black hole which
itself is in equilibrium but whose exterior contains radiation admits such a
horizon.  Inspired by Hamiltonian mechanics, a (quasi-)local definition of
isolated horizon mass is formulated.  Although its definition does not refer
to infinity, this mass takes the standard value in a Reissner--Nordstr\"om
solution.  Furthermore, under certain technical assumptions, the mass of an
isolated horizon is shown to equal the future limit of the Bondi energy.
\end{abstract}

\section{Introduction}
\label{int}

\begin{figure}[t]
  \begin{center}\small
    \psfrag{Ip}{$\scri^+$}
    \psfrag{calM}{$\mathcal{M}$}
    \psfrag{ip}{$i^+$}
    \psfrag{i0}{$\spi$}
    \psfrag{Delta}{$\Delta$}
    \psfrag{M}{$M$}
    \includegraphics[height=6cm]{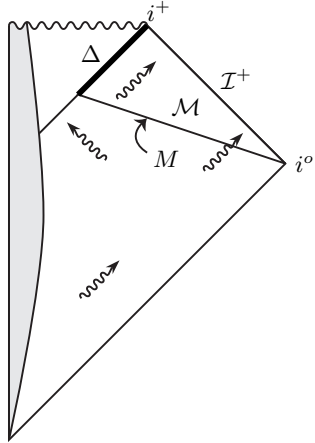}
    \caption{A typical gravitational collapse.  The portion $\Delta$ of the
      horizon at late times is isolated.  The spacetime $\mathcal{M}$ of
      interest is the triangular region bounded by $\Delta$, $\scri^+$ and a
      partial Cauchy slice $M$.}\label{collapse}
  \end{center}
\end{figure}

In their standard form, the zeroth and first laws of black hole mechanics
apply only to the relatively small class of stationary black hole solutions. 
These solutions seem too limited to describe physically realistic situations
since the requirement of stationarity precludes the presence of radiation even
far from the horizon.  As an example, consider the gravitational collapse
depicted in figure \ref{collapse}.  Although the horizon will presumably reach
equilibrium at late times, there will also be gravitational and other
radiation near null infinity.  One would hope the familiar laws of black hole
mechanics continue to apply in such situations.  Recently, using the framework
of {\it isolated horizons\/} \cite{ack,abf}, this expectation has been shown
to be correct: The zeroth and first laws extend to a broad class of spacetimes
containing both radiation and black holes which are, however, isolated from
that radiation.  This contribution focuses on the definition of black hole
mass in this expanded, non-stationary setting and collects the results
presented in both of our talks at the Conference.

The key idea of the isolated horizon framework is to replace the global
construct of an event horizon with a set of boundary conditions applied
locally at the surface of the black hole.  These boundary conditions model a
portion of the horizon which is isolated, i.e., across which there is no flux
of gravitational radiation or matter.  However, since the boundary conditions
are applied {\it only\/} at the horizon, the exterior of an isolated horizon
will generically contain both.  For technical simplicity, we confine our
attention here to non-rotating isolated horizons in Einstein--Maxwell theory.

To formulate the laws of black hole mechanics for isolated horizons, one needs
definitions of the ``extrinsic parameters'' of a black hole --- particularly
its mass $M$ and surface gravity $\kappa$ --- which depend {\it only\/} on the
structure available at the horizon.  In the static context, $M$ is taken to be
the ADM mass and $\kappa$ to be the acceleration at the event horizon of that
static Killing field which is unit at infinity.  Although these parameters are
associated with the black hole, they cannot be constructed solely from the
near-horizon geometry; they are genuinely global concepts.  In the non-static
context, the ADM energy {\it cannot\/} be identified with the mass of the
black hole since it will also include contributions from the energy contained
in radiative fields far from the horizon.  Similarly, without a static Killing
field, the usual prescription for normalizing the null generator of the
horizon is no longer applicable and the usual definition of $\kappa$ therefore
fails.  Drawing motivation from the Hamiltonian formulation of an isolated
horizon, this paper shows how the problem of defining its mass can be
overcome.

The details of our construction, along with a more complete set of references,
may be found in \cite{abf}.

\section{Isolated horizons}
\label{hor}

An isolated horizon is defined by a set of boundary conditions which capture
the essential local structure of a stationary event horizon.  The physical
situation we wish to model is illustrated by the example of figure
\ref{collapse}.  The late stages of the collapse pictured here should describe
a non-dynamical, isolated black hole.  However, a realistic collapse will
generate gravitational radiation which must either be scattered back into the
black hole or radiated to infinity.  Physically, one expects most of the
back-scattered radiation will be absorbed rather quickly and, in the absence
of outside perturbations, the black hole will ``settle down'' to a
steady-state configuration.  Although the presence of radiation elsewhere in
spacetime implies $\cal M$ cannot be stationary, the portion $\Delta$ of the
event horizon at late times will describe an isolated black hole.  The
boundary conditions are designed to represent precisely such situations.  They
are also satisfied at the event horizon of a static, Reissner--Nordstr\"om
black hole.

We are now in a position to state our definition.  For clarity, we shall give
a fairly complete statement of the boundary conditions, even though some of
the details will not be needed here.  A {\sl non-rotating, isolated horizon\/}
is a three-dimensional hypersurface $\Delta$ in spacetime which satisfies the
following four conditions:

\begin{enumerate}
  \renewcommand{\theenumi}{\alph{enumi}}
  \renewcommand{\labelenumi}{(\theenumi) }

\item\label{bcKinem} $\Delta$ is null, topologically $S^2\times \Re$, and
equipped with a preferred foliation by two-spheres $\SD$ transverse to its
null normal $\ell^a$.  We will denote the second null normal to the foliation
by $n^a$ and partially fix the normalizations by setting $\ell^a n_a = -1$ and
requiring the pull-back to $\Delta$ of $n_a$ to be curl-free.

\item\label{bcGeom} $\Delta$ is a non-rotating, non-expanding, future boundary
of spacetime.  Specifically, $\ell^a$ is twist-, shear- and expansion-free and
$n^a$ is twist- and shear-free with strictly negative expansion $\theta_{(n)}$
which is constant on each spherical leaf of the preferred
foliation of $\Delta$.%
\footnote{The spherical symmetry of $\theta_{(n)}$ guarantees the uniqueness
of the preferred foliation.}

\item\label{bcEoM} All equations of motion are satisfied at $\Delta$.

\item\label{bcMax} The flux densities of the electric and magnetic fields are
both constant on each two-sphere $\SD$.

\end{enumerate}
Let us summarize the consequences of the boundary conditions which are
directly relevant here.

First, the notion of an isolated horizon is intrinsically local since it is
defined through a set of boundary conditions.  This makes the definition
well-suited to situations such as that of figure \ref{collapse}.  Although the
boundary conditions imply the surface $\Delta$ is isolated from infalling
matter and radiation, the geometry of the exterior region $\cal M$ remains
largely unconstrained.  Even {\it at\/} $\Delta$, the {\it outgoing\/}
radiative modes (i.e., those flowing along $\Delta$) are undetermined.

Second, a non-rotating isolated horizon in Einstein--Maxwell theory is
characterized by three parameters; its radius $\rad$, electric charge $\QD$
and magnetic monopole moment $\PD$.  Much of the intrinsic structure of
$\Delta$ depends only on this remarkably small set of parameters.  In
particular, the Newman--Penrose components $\Psi_2$ and $\phi_1$ \cite{pr1},
describing respectively the ``Coulombic parts'' of the gravitational and
electromagnetic fields at $\Delta$, are functions of these variables alone.

Third, the boundary conditions guarantee all three parameters are constant in
time.  The radius is time-independent due to the vanishing expansion of
$\Delta$ and the imposition of the Maxwell equations at $\Delta$ implies the
same for the electromagnetic charges.  It is important to note that this
time-independence is a result of {\it boundary conditions\/} and not of
dynamics.  Consequently, this result holds in {\it any\/} history compatible
with the boundary conditions and not just ``on-shell.''

Before moving on to our discussion of mass, let us briefly consider the
surface gravity.  As discussed in the Introduction, the standard formulation
of black hole mechanics defines surface gravity as the acceleration of the
appropriately normalized null generator of the horizon.  This normalization is
possible since the horizon-generating Killing field can be fixed to an unit
time translation at infinity.  However, a generic isolated horizon spacetime
possesses no such Killing field.  To overcome this problem, we must provide a
means of fixing the normalization of $\ell^a$ on {\it any\/} isolated horizon;
then $\kappa$ may again be defined as the acceleration of the properly
normalized $\ell^a$.  Recall that $\ell^a$ itself is null and free of shear,
expansion and twist, whence we cannot normalize it through its intrinsic
properties.  However, since $\theta_{(n)}$ is strictly negative by condition
(\ref{bcGeom}), we {\it can\/} normalize $n^a$ by setting its expansion to
some given value.  This in turn will fix the normalization of $\ell^a$ since
$\ell^a n_a = -1$.

Fortunately, a preferred value of $\theta_{(n)}$ exists.  In a
Reissner--Nordstr\"om solution, with $\ell^a$ taken to be the restriction to
$\Delta$ of the properly normalized static Killing field, one finds
$\theta_{(n)} = -2/\rad$.  Since we want to include this family of solutions
in our analysis, let us require $n^a$ (and hence $\ell^a$) {\it always\/} be
normalized such that $\theta_{(n)} = -2/\rad$.  This convention, which is {\it
not\/} a consequence of the boundary conditions, provides a definition of
surface gravity for a generic isolated horizon: The value of $\kappa$ is
simply the acceleration of the properly normalized $\ell^a$.

\section{The Hamiltonian and black hole mass}
\label{ham}

A key question faced by any set of boundary conditions is whether they allow
the formulation of a well-defined action principle.  For boundary conditions
enforcing asymptotic flatness in spacetimes with no interior boundaries, it is
well known that an action principle can be found by adding a boundary term at
infinity to the standard bulk action.  The situation is similar when one
allows an isolated horizon at an interior boundary.  A well-defined action
principle can then be obtained by also including a boundary term at $\Delta$
\cite{abck}.  However, as we will see now, there is some subtlety in using
this action principle to formulate a Hamiltonian description of an isolated
horizon system.

In passing to the Hamiltonian framework, one performs a Legendre transform of
the action.  To do so, let us foliate the spacetime $\cal M$ with partial
Cauchy surfaces $M$ which extend to spatial infinity and whose inner
boundaries are the preferred two spheres $\SD$.  Next, fix a smooth vector
field $t^a$, transverse to the leaves $M$, which tends to a $\ell^a$ at the
horizon%
\footnote{Here, $\ell^a$ denotes the null generator of $\Delta$ normalized
according to the prescription given at the end of the previous section. 
Physically, this vector field defines the ``rest frame'' of the black hole.}
and to an unit time translation orthogonal to $M$ near infinity.  Then, after
the Legendre transform, the Hamiltonian corresponding to evolution along $t^a$
can be written in the form
\begin{equation}\label{Htot}
  \Ht = \int_M (\mbox{constraints}) + \HI - \HD.
\end{equation}
The quantities $\HI$ and $\HD$ represent the surface terms in the Hamiltonian
at infinity and $\Delta$ respectively.  The term at infinity is the usual one
arising from the asymptotically flat boundary conditions imposed there.  It
can be expressed in terms of components of the Weyl curvature as
\begin{equation}\label{Hinf}
  \HI = \lim_{R \to \infty} \oint_{S_R^{}} \left[ 
    -\frac{R}{4\pi G} \Psi_2 \right] \, {}^2\!\epsilon.
\end{equation}
On any solution to the field equations, this surface term will equal the ADM
energy of spacetime.

On the other hand, due to an ambiguity in the action caused by the variational
principle, the horizon boundary term $\HD$ in the Hamiltonian is initially an
{\it arbitrary\/} function of the parameters $\rad$, $\QD$ and $\PD$.  Since
every allowable history must satisfy our boundary conditions, the variation of
these parameters must be time-independent.  However, the action principle
allows only variations which vanish at the initial and final time slices.  It
follows that $\delta\rad$, $\delta\QD$ and $\delta\PD$ vanish everywhere in
the Lagrangian formalism.  Therefore, one may add {\it any\/} function of
these parameters to the action and still have a well-posed variational
principle.  As a result of this freedom, $\HD$ will not be fixed by the
Lagrangian.

The ambiguity in $\HD$ is resolved within the Hamiltonian formalism.  The
reason for this lies in subtle differences between the Lagrangian and
Hamiltonian variations.  In the Hamiltonian framework, the phase space
consists of fields on a fixed spacelike three-manifold $M$ which describe
horizons of arbitrary radius and charges.  Consequently, there {\it are\/}
tangent vectors $\delta$ to phase space which change the values of $\rad$,
$\QD$ and $\PD$.  Requiring the consistency of Hamilton's equations under such
variations in phase space determines the Hamiltonian {\it uniquely\/}.  The
resulting horizon surface term in the Hamiltonian is
\begin{equation}\label{Hhor}
  \HD = \oint_{\SD} \left[ -\frac{\rad}{4\pi G} \, \Psi_2 + 
      \frac{\QD - i\PD}{2\pi\rad} \, \phi_1 \right] \, {}^2\!\epsilon,
\end{equation}
where the Newman--Penrose components $\Psi_2$ and $\phi_1$ are determined in
terms of $\rad$, $\QD$ and $\PD$ by the boundary conditions.

Remarkably, in a static Reissner--Nordstr\"om solution, one finds the horizon
surface term (\ref{Hhor}) is exactly equal to that at infinity (\ref{Hinf}). 
Consequently, the full Hamiltonian vanishes in a static solution.  This
feature is not accidental: There is a general argument from symplectic
geometry which requires the vanishing of $\Ht$ on a stationary solution.

Now let us turn to the definition of black hole mass.  In physical theories,
energy is defined as the on-shell value of the Hamiltonian generating an
appropriate time translation.  In general relativity, the bulk term in the
Hamiltonian consists solely of constraints and therefore vanishes on-shell. 
Thus, the energy of a system is given by the surface terms in its Hamiltonian. 
For example, in asymptotically flat spacetime, the Hamiltonian surface term at
infinity is precisely equal to the ADM energy when the constraints are
satisfied.  In analogy, we {\it define\/} the mass of an isolated horizon to
be the horizon surface term in the Hamiltonian:
\begin{equation}\label{Mhor}
  \MD := \HD = \oint_{\SD} \left[ -\frac{\rad}{4\pi G} \, \Psi_2 + 
    \frac{\QD - i\PD}{2\pi\rad} \, \phi_1 \right] \, {}^2\!\epsilon.
\end{equation}
This formula for black hole mass resembles the expression (\ref{Hinf}) for ADM
mass, but includes an additional contribution from the electromagnetic field. 
Without this contribution, $\MD$ would {\it not\/} give the correct mass for a
Reissner--Nordstr\"om black hole.  For a generic isolated horizon, the
expression (\ref{Mhor}) for $\MD$ can be regarded as ``the mass of the black
hole together with its Coulombic hair.''  That is, it includes the energy
associated with static fields emanating from $\Delta$, but {\it not\/}
contributions due to radiative excitations outside $\Delta$.

\section{Radiative energy and black hole mass}
\label{rad}

As discussed in the Introduction, the ADM energy cannot be used to measure the
mass of a non-stationary black hole since it also includes contributions from
the radiative modes of the fields.  This raises the question of whether there
exists a precise relation among the ADM energy, the black hole mass and the
radiative energy.  One would expect such a relation to be simply
\begin{equation}\label{Esum}
  (\mbox{ADM energy}) = (\mbox{black hole mass}) + (\mbox{radiative energy}).
\end{equation}
We will now show that, under certain technical assumptions, this relation
holds within the framework of isolated horizons.  This result is on a somewhat
different footing from the rest of the calculations presented here due to the
additional assumptions needed to complete the proof.  Nevertheless, from a
physical point of view, the relation (\ref{Esum}) serves to strengthen our
intuition regarding isolated horizon mass and provides further justification
for our definition.

Consider an isolated horizon, such as $\Delta$ in figure \ref{collapse}, which
extends to future time-like infinity $i^+$.  Since no radiation can cross
$\Delta$, all the radiation in $\cal M$ must ``register on'' future null
infinity $\scri^+$.  It is therefore not surprising that the radiative data
can be encoded in fields {\it on\/} $\scri^+$ \cite{aa}.  Furthermore, these
fields on $\scri^+$ admit a phase space structure \cite{as}.  That is, there
exists a symplectic structure $\ONI$ on the space of data at null infinity and
a Hamiltonian $\HNI$ which generates time evolution in that space.  Moreover,
on any solution to the field equations, the value of $\HNI$ is precisely equal
to the total radiative energy.  This fact is the key to the proof of
(\ref{Esum}) for isolated horizon systems.

We now have two phase spaces; the isolated horizon phase space $(\Omega, \Ht)$
discussed in the previous section and the asymptotic phase space $(\ONI,
\HNI)$ introduced here.  Both of these describe the radiative modes of the
fields in $\cal M$, and hence it is reasonable to expect they should also be
``equivalent'' in some appropriate sense.  This expectation turns out to be
correct.  There is a natural map from the isolated horizon phase space to the
phase space at null infinity such that $\Omega \mapsto \ONI$.  Using this
fact, we can apply Hamilton's equations in each phase space to find
\begin{equation}\label{Heq}
  \delta \Ht = \Omega(\delta, X_H) = \ONI(\delta, X_H) = \delta \HNI.
\end{equation}
It follows that the isolated horizon Hamiltonian and its counterpart at null
infinity differ at most by a constant.  In the previous section, however, we
showed the isolated horizon Hamiltonian $\Ht$ vanishes on a
Reissner--Nordstr\"om solution.  Furthermore, these static solutions contain
no radiation, whence the Hamiltonian at $\scri^+$ must also vanish. 
Therefore, the ``constant of integration'' for (\ref{Heq}) is zero and the two
Hamiltonians are {\it equal\/} on any solution to the equations of motion. 
Since $\Ht$ equals the difference between the ADM energy and the black hole
mass by (\ref{Htot}) and $\HNI$ equals the radiative energy in spacetime, the
relation (\ref{Esum}) follows.

Therefore, {\it the mass of an isolated horizon is equal to the ADM energy of
spacetime minus the energy contained in radiation outside the horizon\/}.  It
is well known that the difference of the ADM energy and the flux of energy
through $\scri^+$ is the future limit of the Bondi energy.  Hence, isolated
horizon mass is the future limit of the Bondi energy.  Equivalently, $\MD$ can
be thought of as the mass remaining in spacetime after all radiation has
escaped to infinity.

\section{Summary}
\label{sum}

We have presented the definition of a non-rotating isolated horizon.  Within
the surrounding framework, we have shown how the definitions of mass (and
surface gravity) can be extended to a non-static context.  Although it has not
been discussed here, these definitions enable one to formulate and prove the
zeroth and first laws of black hole mechanics for generic isolated horizons
\cite{abf}.

The mass $\MD$ is given by the horizon surface term in the {\it unique\/}
consistent Hamiltonian for an isolated horizon system.  It is manifestly
(quasi-)local to the horizon and correctly reproduces the mass of
Reissner--Nordstr\"om black holes.  Furthermore, given certain technical
assumptions, one can show the ADM energy of a spacetime containing a single
isolated horizon is precisely the sum of $\MD$ and the total energy contained
in the radiative modes of the fields.  In this case, $\MD$ is equal to the
future limit of the Bondi energy.

\bigskip\noindent{\bf Acknowledgements\/} \quad We would like to express our
gratitude to Abhay Ashtekar for his collaboration in the work reported here. 
The authors were supported in part by NSF grants PHY95-14240 and INT97-22514
and by the Eberly research funds of the Pennsylvania State University.

\end{document}